\documentclass[english,aps,floats,twocolumn,showpacs,nofootinbib]{revtex4}
\usepackage{pslatex}
\usepackage[T1]{fontenc}
\usepackage[latin1]{inputenc}
\usepackage{graphicx}
\usepackage{epsfig}

\usepackage{calc}
\usepackage{ifthen}

{
{
{
\newcommand{\bea}{\begin{eqnarray}}
\newcommand{\eea}{\end{eqnarray}}

\newcommand{\nc}{\newcommand}
\nc{\renc}{\renewcommand}
\nc{\eqs}[2]{\mbox{Eqs.~(\ref{#1},\,\ref{#2})}}
\nc{\eq}[1]{\mbox{Eq.~(\ref{#1})}}
\nc{\figs}[2]{\mbox{Figs.~(\ref{#1},\,\ref{#2})}}
\nc{\fig}[1]{\mbox{Fig~.(\ref{#1})}}
\nc{\be}[1]{\begin{equation} \mbox{$\label{#1}$}}
\nc{\ee}{\vspace{0.1cm}\end{equation}}

\newcommand{\bean}{\begin{eqnarray*}}
\newcommand{\eean}{\end{eqnarray*}}

\def\GeV{{\rm \ GeV}}
\def\MeV{{\rm \ MeV}}

\def\TeV{{\rm \ TeV}}

\def\lae{\;^{<}_{\sim} \;}

\begin{document}
\title{Enhanced Dark Matter Annihilation Rate for Positron and Electron Excesses from Q-ball Decay}
\author{John McDonald}
\email{j.mcdonald@lancaster.ac.uk}
\affiliation{Cosmology and Astroparticle Physics Group, University of
Lancaster, Lancaster LA1 4YB, UK}
\begin{abstract}

        We show that Q-ball decay in Affleck-Dine baryogenesis models can account for dark matter when the annihilation cross-section is sufficiently enhanced to explain the positron and electron excesses observed by PAMELA, ATIC and PPB-BETS. For Affleck-Dine baryogenesis along a $d = 6$ flat direction, the reheating temperature is approximately 30 GeV and the 
Q-ball decay temperature is in the range 10-100 MeV. The LSPs produced by Q-ball decay
annihilate down to the observed dark matter density if the cross-section is enhanced
by a factor $\sim 10^3$ relative to the thermal relic cross-section.

\end{abstract}
\maketitle
\section{Introduction}

     Recent observations of a cosmic-ray positron excess by PAMELA \cite{pam} and an electron excess by ATIC \cite{atic} and PPB-BETS \cite{bets} are consistent with dark matter of mass $\sim 1 \TeV$
which annihilates primarily to leptons and which has an annihilation cross-section enhanced relative to the thermal relic cross-section by a factor $ \sim 10^3$ \cite{enhance}. Explaining this 
enhancement factor requires new physics beyond standard thermal relic dark matter. 
Two main possibilities have been considered: (i) Sommerfeld enhancement of the annihilation
cross-section \cite{enhance,se,se2,se3} and (ii) non-thermal production of dark matter via out-of-equilibrium decay of a       
heavier particle \cite{nontherm}. Sommerfeld enhancement uses a light boson to mediate a long-range attractive force between the dark matter particles. This requires a hierarchy between the light boson mass $m_{B}$ and the dark matter particle mass $m_{\chi}$ of the form $m_{B} \approx \alpha_{B} m_{\chi}$, where $\alpha_{B}$ is the fine-structure constant of the interaction.  The mass is typically $\sim$ 10 GeV for dark matter particle $\sim$ 1 TeV. 
In the context of gravity-mediated SUSY breaking this may be difficult to implement, as scalars have SUSY breaking masses which are typically several hundred GeV. Possibility (ii) is also not so easy: in order to 
enhance the 
cross-section the heavy particles must decay well after the dark matter particles have frozen-out, at temperatures $\lae 10 \GeV$. This requires very small couplings to have a sufficiently slow decay rate, e.g. if the decay rate is $\Gamma_{d} 
\approx \lambda_{X}^2 m_{X}/4 \pi$, where $X$ is the heavy decaying particle, then $\lambda_{X} \lae 10^{-9}$ is necessary in order to have $\Gamma_{d} < H$ at $T = 10 \GeV$. So solution (i) or (ii) will tend to undermine the naturalness of SUSY dark matter models. 

      In this note we point out that an alternative source of non-thermal dark matter exists in the MSSM and its extensions, namely Q-ball decay \cite{jq1,jq2,jq3,fujii,fujii2}. Q-balls are a natural possibility in SUSY models, forming from fragmentation of flat direction condensates \cite{ks}. When combined with Affleck-Dine baryogenesis \cite{ad}, they can provide a common origin for the baryon asymmetry and dark matter, with the scalars of the Q-ball decaying to the lightest SUSY particles (LSPs) and baryons \cite{jq1,jq2}. Generally the LSP density produced by Q-ball decay is initially too large, requiring subsequent annihilation. In the following we will show that the required annihilation cross-section can naturally be consistent with an enhancement factor $\sim 10^{3}$.

\section{Enhancement factor from Q-ball decay} 

       In the standard thermal relic dark matter scenario, the value of the annihilation cross-section 
times relative velocity necessary to produce the observed dark matter density is $<\sigma v>_{th} 
\approx 3 \times 10^{-26} {\rm cm^3 /s} \equiv 2.6 \times 10^{-9} \GeV^{-2}$. This is too small 
to account for the positron and electron excesses, which require a present annihilation rate which is enhanced 
by a factor $\sim 10^3$ over that expected for a smooth distribution of galactic dark matter with standard thermal relic value for $<\sigma v>$. The annihilation rate could be enhanced by a boost factor from galactic substructure \cite{subs}, although it has been argued that small boost factors $\sim 2-3$ are likely \cite{subs2}. In the following we will consider the enhancement to come entirely from the annihilation cross-section, with $<\sigma v> = B_{ann} <\sigma v>_{th}$.

     Due to their relatively small number density, the self-annihilation rate of LSPs will be very small compared to their scattering rate from thermal background particles. As a result, they will slow by scattering before annihilating. Therefore we will assume that the LSPs from Q-ball decay are non-relativistic when they annihilate. 
The freeze-out number density after Q-ball decay at $T_{d}$ is given by
\be{a1} n(T_{d})  = \frac{H(T_{d})}{<\sigma v>}      ~,\ee 
where $H$ is the expansion rate. 
We will consider the case where $<\sigma v>$ is a constant, which is true for scalars and Dirac fermions. Assuming that the Q-balls decay during radiation-domination, the 
LSP density at present following from \eq{a1} is then
\be{a2}  \Omega = \frac{1}{B_{ann}} \left(\frac{4 \pi^3}{45} \right)^{1/2} \frac{g(T_{\gamma})}{g(T_{d})^{1/2}} \frac{T_{\gamma}^{3}}{T_{d} M_{Pl}}  \frac{m_{\chi}}{\rho_{c} <\sigma v>_{th} }    ~,\ee
where $T_{\gamma} = 2.4 \times 10^{-13} \GeV$ is the present photon temperature, $g(T)$ is the effective number of degrees of freedom and $\rho_{c} = 8.1 \times 10^{-47} h^2 \GeV^4$ is the critical density. This gives
\be{a3}   \Omega h^2 = 0.55 \left( \frac{10^{3}}{B_{ann}}\right)  
\left(\frac{10.75}{g(T_{d})}\right)^{1/2} 
\left( \frac{10 \MeV}{T_{d}}\right)
\left( \frac{m_{\chi}}{1 \TeV}\right)    ~.\ee  
Observation requires $\Omega h^{2} = 0.1131 \pm 0.0034$ \cite{komatsu}. 
The enhancement factor as a function of Q-ball decay temperature is therefore given by 
\be{a4} B_{ann} = 10^{3}\left(\frac{10.75}{g(T_{d})}\right)^{1/2} 
 \left( \frac{m_{\chi}}{1 \TeV}\right)
 \left( \frac{50 \MeV}{T_{d}}\right)    ~.\ee  
Thus the enhancement factor is the right order of magnitude when dark matter comes from Q-ball decay and the decay temperature is in the range $T_{d} \sim 10-100 \MeV$. We next show that such decay temperatures can arise naturally in the context of Affleck-Dine baryogenesis in gravity-mediated SUSY breaking.    

\section{Q-ball formation and decay} 

     The physics of Q-ball formation and decay in gravity-mediated SUSY breaking was first discussed in \cite{jq1,jq2}, with the formation of Q-balls confirmed numerically in \cite{kk}. The possibility of dark matter from Q-ball decay was introduced in \cite{jq2,jq3} and further analysed in \cite{fujii}. In \cite{fujii2} it was noted that non-thermal dark matter produced by Q-ball decay could produce a potentially observable positron signal in the case of Higgsino- and Wino-like dark matter. We summarize the relevant physics below. 

       A flat direction of dimension $d$ is described by an effective superpotential 
\be{e1}    W = \frac{\lambda \Phi^{d}}{d!M^{d-3}}     ~,\ee
where $M = M_{Pl}/\sqrt{8 \pi}$. $\lambda$ is of order 1 in the case where the physical strength of the coupling is determined by $M$. The scalar potential including soft-SUSY breaking terms and Hubble corrections is then 
\be{e2} V(\Phi) = (m^2 - cH^2) |\Phi|^2 + \frac{\lambda^2 |\Phi|^{2(d-1)}}{(d-1)!^2 M^{2(d-3)} } + ({\rm A-terms}) 
~,\ee 
where $m$ is the SUSY-breaking scalar mass and $|c|$ is of order 1.  
Oscillations of the condensate begin once the expansion rate $H = H_{osc} \approx m/c^{1/2}$, with initial amplitude 
\be{e3} |\Phi|_{osc}^2 \approx \kappa_{d} \left(m^2 M^{2(d-3)}\right)^{1/(d-2)}    ~,\ee
where 
\be{e4} \kappa_{d} = \left( \frac{\left(d-1\right)!^{2}}{\lambda^2 \left(d-1\right)} 
\right)^{1/(d-2)}    ~.\ee
$\Phi$ oscillations begin during the inflaton oscillation dominated era. We assume that the baryon asymmetry is generated by Affleck-Dine baryogenesis. If the baryon asymmetry induced in the condensate by the A-term is maximal, then the initial baryon number density is $n_{B\;osc} \approx \rho_{\Phi\;osc}/m$, where $\rho_{\Phi\;osc} = m^2 |\Phi|_{osc}^2$. After scaling down to the present temperature, the baryon-to-entropy ratio following from \eq{e3} is given by 
\be{e5} \eta_{B} \approx \frac{\kappa_{d} T_{R}}{4 M} \left(\frac{M}{m}\right)^{\frac{d-4}{d-2}} 
~.\ee
Therefore the reheating temperature as a function of $\eta_{B}$ is  
\be{e6} T_{R} \approx \frac{4 \eta_{B} M}{\kappa_{d}} \left(\frac{m}{M}\right)^{\frac{d-4}{d-2}}   ~.\ee 
In Table 1 we give the reheating temperature necessary to account for the observed baryon asymmetry, $\eta_{B} \approx 1.5 \times 10^{-10}$, for a range of values of $d$.

\begin{table}[h]
 \begin{center}
 \begin{tabular}{|c|c|}
	\hline   $d$ &  $\kappa_{d}T_{R}$ \\ 
	\hline   4 & $1.4 \times 10^{9} \GeV$ \\	
      \hline   5 & $1.1 \times 10^{4} \left(\frac{m}{1 \TeV}\right)^{1/3} \GeV$ \\	
      \hline   6 & $29 \left(\frac{m}{1 \TeV}\right)^{1/2} \GeV$  \\	
      \hline   7 & $0.85 \left(\frac{m}{1 \TeV}\right)^{3/5} \GeV$  \\	
      \hline   8 & $0.08 \left(\frac{m}{1 \TeV}\right)^{2/3} \GeV$  \\	
      \hline 
 \end{tabular}
 \caption{\footnotesize{Reheating temperature as a function of $d$ for successful Affleck-Dine baryogenesis}}  
 \end{center}
 \end{table}

     Due to quantum corrections, the flat direction condensate is unstable with respect to spatial perturbations and fragments to form Q-balls. Once $cH^2 \ll m^2$, the potential is given by
\be{e7} V(\Phi) \approx m^2(1 + K \ln (|\Phi|^2/\Lambda^2))|\Phi|^2    ~,\ee
where $\Lambda$ is a renormalization scale. 
$K$ is mostly due to gaugino loops, with $|K|$ in the range 0.01 to 0.1. The perturbations become non-linear and the condensate fragments when 
\be{e8}  H_{i} \approx \frac{2 |K| m}{\alpha}      ~,\ee
where $\alpha = -\ln \left(\frac{\delta \phi_{0}}{\phi_{0}}\right)$ and $\delta \phi_{0}/\phi_{0}$ is the initial perturbation of the field at $H_{osc}$. 
Spatial perturbations of wavenumber $k^2/a^2 \approx 2 |K|m^{2}$ experience the largest growth at a given time. 
The radius of the lumps at fragmentation is then given by 
\be{e9} \lambda_{i} \approx \frac{\pi}{(2 |K| m^2 )^{1/2}}   ~.\ee
Assuming that the baryon asymmetry also comes from the condensate, the baryon density when $T > T_{R}$ is 
\be{e10} n_{B}(H) = \frac{\eta_{B}}{2 \pi} \frac{H^2 M_{Pl}^2}{T_{R}}   ~.\ee
Therefore the baryon number charge of the Q-ball of radius $\lambda_{i}$ at fragmentation is
\be{e11} Q = \frac{4 \pi \lambda_{i}^3 n_{B}(H_{i})}{3} \approx \frac{2 \sqrt{2} \pi^3}{3} \frac{\eta_{B} |K|^{1/2} M_{Pl}^2}{\alpha^2 m T_{R}}   ~.\ee
The simplest origin for the perturbations $\delta \phi_{0}$ are adiabatic fluctuations of the inflaton which have re-entered the horizon before $H \approx H_{osc}$, which implies that 
$\delta \phi_{0}/\phi_{0} \sim 10^{-5}$ and $\alpha \approx 12$. (This is somewhat smaller than the original estimate of 40 given in \cite{jq1,jq2}). The 
mean charge of the Q-balls is therefore 
\be{e12}  Q \approx 2.9 \times 10^{22} |K|^{1/2} \left( \frac{1 \TeV}{m} \right) \left(\frac{12}{\alpha}\right)^2 \left( \frac{\eta_{B}}{10^{-10}} \right) \left( \frac{100 \GeV}{T_{R}} \right)    ~.\ee

    Q-balls have a long lifetime because the flat direction scalars can 
decay only close to the surface, resulting in a decay rate proportional to the area of the Q-ball. For the case of $\Phi$ decaying to a final state with a fermion, Pauli blocking of the decay within the volume leads to an upper bound on the decay rate \cite{cole}
\be{e13} \left(\frac{dQ}{dt}\right)_{fermion} \leq \frac{\omega^3 A}{192 \pi^2}     ~,\ee 
where $\omega$ is the angular frequency of the Q-ball solution. For gravity-mediated Q-balls $\omega \approx m$ \cite{jq2}.   
We will assume the upper limit is saturated in the following. In the case where $\Phi$ decays to scalars 
there is no Pauli blocking, but decay still occurs close to the suface of the Q-ball since the large $\Phi$ field inside the Q-ball gives a large mass to fields coupling to $\Phi$, kinematically suppressing $\Phi$ decay. The scalar decay rate may be enhanced by a factor $f_{S}$ relative to the fermion case, where $f_{S} \approx 10^3$ was estimated in \cite{jq2}. The 
decay temperature of the Q-ball is then 
\be{e14a} T_{d} \approx \left(\frac{f_{S} \omega^3 R^2 M_{Pl}}{48 \pi k_{T} Q} \right)^{1/2} ~,\ee
where $k_{T} = (4 \pi^3 g(T_{d})/45)^{1/2}$ and $R \approx (|K|^{1/2} m)^{-1}$ is the radius of the Q-ball \cite{jq2}. Therefore
\be{e14} T_{d} \approx 0.38 \left(\frac{10.75}{g(T_{d})}\right)^{1/4} \left(\frac{f_{S}}{|K|}\right)^{1/2} \left( \frac{m}{1 \TeV} \right)^{1/2} 
\left( \frac{10^{20}}{Q} \right)^{1/2} \GeV    ~.\ee

     Given the dimension $d$ of the flat direction, the reheating temperature necessary to generate the observed baryon asymmetry can be found from \eq{e6} and so the Q-ball charge and the temperature at which the Q-balls decay can be obtained from 
\eq{e12} and \eq{e14} respectively. For a $d = 6$ flat direction, we find (assuming $\eta_{B} = 1.5 \times 10^{-10}$)  
\be{e15} Q \approx 4.7 \times 10^{22} \; \kappa_{6} \left( \frac{|K|}{0.1} \right)^{1/2} \left(\frac{12}{\alpha}\right)^2 \left( \frac{1 \TeV}{m} \right)^{3/2}  ~\ee
and 
\be{e16} T_{d} \approx 55 \MeV 
\left(\frac{f_{S}}{\kappa_6}\right)^{1/2} \left( \frac{0.1}{|K|} \right)^{3/4} \left(\frac{10.75}{g(T_{d})}\right)^{1/4}  \left(\frac{\alpha}{12}\right)  \left( \frac{m}{1 \TeV} \right)^{5/4}   ~.\ee 
The corresponding reheating temperature assuming $\kappa_6 \approx 1$ is $T_{R} \approx 30 $GeV, so Q-ball decay occurs during the radiation-dominated phase following reheating.  
Thus $T_{d} \sim 10-100$ MeV is natural in Affleck-Dine baryogenesis for the case of Q-ball decay to a final state with a fermion. From \eq{a4} this implies 
an enhancement factor $\sim 10^3$ if dark matter originates from Q-ball decay. Therefore Q-ball decay to dark matter can easily account for the dark matter density when the annihilation rate is enhanced by the factor necessary to explain the observed positron and electron excesses.

\section{Conclusions}

       We have shown that in the case of $d = 6$ 
Affleck-Dine baryogenesis, the Q-ball decay temperature is in the correct range $10-100 \MeV$ necessary to produce the observed dark matter density when the annihilation rate is enhanced by a factor $ \sim 10^3$. Since Q-ball formation and decay is a natural feature of the MSSM and its extensions, this provides a way to account for the dark matter density in the presence of an enhanced annihilation rate without introducing either light scalars for Sommerfeld enhancement or heavy decaying particles with unusually long lifetimes, both of which 
tend to undermine the naturalness of SUSY dark matter models. 

       Observations find a positron and electron excess without an anti-proton excess. 
In order to explain this, a model is required where the dark matter particles annihilate primarily to lepton final states. An interesting possibility is right-handed (RH) sneutrino dark matter. In \cite{all1} the MSSM was extended by a $U(1)_{B-L}$  
gauge group and the addition of Higgs fields $H_{1}^{'}$ and $H_{2}^{'}$ which break the $U(1)_{B-L}$ at the TeV scale\footnote{In \cite{all2} a related model was suggested in which the dark matter is a $U(1)_{B-L}$ neutralino. Here we will focus on the scalar dark matter case.}. The LSP is a RH sneutrino, which annihilates to the lightest $U(1)_{B-L}$ Higgs $\phi$. This subsequently decays primarily to $\tau^{+} \tau^{-}$ via $U(1)_{B-L}$ bosons at 1-loop. The LSP mass is $\sim 1 $TeV while the mass of $\phi$ must be $\sim 10$ GeV in order to Sommerfeld
enhance the RH sneutrino annihilations. While not impossible, such a large hierarchy in scalar masses is an awkward feature in an otherwise attractive model.  Replacing Sommerfeld enhancement by Q-ball decay can avoid this hierarchy. A $d=6$ $SU(3)_{c} \times SU(2)_{L} \times U(1)_{Y} \times U(1)_{B-L}$ flat direction is defined by the monomial $d^{c}d^{c}d^{c}LLH_{1}^{'}$, with $U(1)_{B-L}$ charges assigned according to the minimal model of \cite{all1}. Provided that the 
Yukawa couplings of the superfields are not too large \cite{asko}, Affleck-Dine baryogenesis with Q-ball formation can occur along this direction. These Q-balls will decay at the right temperature to produce the dark matter density when the annihilation rate has the required $\sim 10^{3}$ enhancement factor. This assumes that the decay of the flat direction scalars is to a final state containing a fermion, which will be true if the mass difference between $\Phi$ and the LSP is less than the mass of the Higgs scalars, since in this case at least one fermion must appear in the final state.

    For a complete model we need a dark matter candidate with an annihilation cross-section boosted by a factor $B_{ann}$ 
over the thermal relic annihilation cross-section. For scalar dark matter this is not possible with a simple perturbative annihilation process. For the RH sneutrino, the annihilation is primarily to $\phi$ pairs. If the coupling is 
$\lambda  \phi^2 |\tilde{N}|^2$ then the annihilation cross-section times relative velocity is 
$ <\sigma v_{rel}>   = \lambda^2/(64 \pi m_{\tilde{N}}^2)$. 
To have a boost factor $B_{ann}$ then requires that $\lambda \approx 23 (m_{\tilde{N}})/(1 \TeV) \left(B_{ann}/10^3\right)^{1/2}$. Therefore a perturbative annihilation process cannot produce a large enough boost factor. There are two ways that this 
might be overcome. The simplest is to consider a strongly coupled $U(1)_{B-L}$. This can produce a non-perturbative sneutrino annihilation cross-section to $\phi$ pairs, since $\lambda \propto g_{B-L}^2$ \cite{all1}. Since $U(1)_{B-L}$ is broken at the TeV scale, this will not affect low energy Standard Model physics. In this case the only constraint on the boost factor is from unitarity, which imposes an upper bound on $<\sigma v_{rel}>$ which is equivalent to $m_{\tilde{N}} \lae 3 (10^3/B_{ann})^{1/2} \TeV$ \cite{unitarity}.  The model will be otherwise similar to the Sommerfeld enhanced scenario of \cite{all1}, with $\phi$ decaying primarily to $\tau^{+}\tau^{-}$ if $m_{\phi} < 2 m_{t} \approx 350 \GeV$. (For larger  $\phi$ masses decay to t-quark pairs becomes dominant.) However, there is no need to have a very small $\phi$ mass relative to the LSP mass in this scenario. Breit-Wigner resonant annihilation may also allow for an enhanced annihilation cross-section \cite{bw1,bw2}. 
This occurs if $m_{\tilde{N}} \approx m_{\Phi}/2$, where $\Phi$ is the heavier of the real mass eigenstate $U(1)_{B-L}$ Higgs bosons \cite{all1}. Although this requires a strong coincidence between the $\tilde{N}$ and $\Phi$ mass, it does not require a small $\phi$ mass relative the the LSP mass.

   We finally note that in this class of model both the baryon asymmetry and dark matter originate from a flat direction condensate. As a result, baryon and dark matter isocurvature perturbations can be generated \cite{iso1,iso2,iso3}. We will return to this possibility in future work.

\section*{Acknowledgement}
This work was supported by the European Union through the Marie Curie Research and Training Network "UniverseNet" (MRTN-CT-2006-035863).

\end{document}